\documentclass[10pt,prd,aps,twocolumn,preprintnumbers,showpacs,nofootinbib,superscriptaddress,notitlepage]{revtex4-1}

\usepackage{mathrsfs}
\usepackage{amsfonts}
\usepackage{amsmath}
\usepackage{slashed}
\usepackage{array}
\usepackage{verbatim}
\usepackage{epsfig}
\usepackage{graphicx}
\usepackage{color}
\usepackage[dvipsnames]{xcolor}
\usepackage[normalem]{ulem}

\begin{document}

\title{Redshift Evolution of the HII Galaxy $L$--$\sigma$ Relation: Gaussian Process Analysis and Cosmological Implications}

\author{Jiaze Gao}
\affiliation{Institute of Theoretical Physics, School of Physics, Dalian University of Technology\\ Dalian 116024, People’s Republic of China}

\author{Yun Chen}
\email{chenyun@bao.ac.cn}
\affiliation{National Astronomical Observatories, Chinese Academy of Sciences\\
Beijing 100101, China}
\affiliation{College of Astronomy and Space Sciences, University of Chinese Academy of Sciences\\
Beijing, 100049, China}

\author{Lixin Xu}
\email{lxxu@dlut.edu.cn}
\affiliation{Institute of Theoretical Physics, School of Physics, Dalian University of Technology\\ Dalian 116024, People’s Republic of China}

\date{\today}

\footnotetext{\noindent \copyright 2026 American Physical Society. This is the accepted manuscript of the following article: Jiaze Gao, Yun Chen, Lixin Xu, "Redshift Evolution of the HII Galaxy $L$--$\sigma$ Relation: Gaussian Process Analysis and Cosmological Implications", Phys. Rev. D 113, 103527 (2026). The final published version is available at \url{https://doi.org/10.1103/6tx3-prh3}. This manuscript version is posted with permission for non-commercial scholarly use.}

\begin{abstract}
The empirical correlation between the H$\beta$ luminosity ($L$) and the ionized gas velocity dispersion ($\sigma$) in HII starburst galaxies (HIIGs) provides a foundation for using them as cosmological standard candles. A key unresolved issue is whether this $L$--$\sigma$ relation changes with redshift, which would impact its application at high redshifts. We test for possible evolution using cosmology-independent distance estimates up to $z \sim 1.8$, obtained from Gaussian Process regression of the Pantheon+ Type Ia supernovae Hubble diagram. These distances allow us to compare the standard $L$--$\sigma$ relation with three redshift-dependent extensions through Bayesian model comparison. We find that a logarithmic redshift correction is statistically preferred when the intrinsic dispersion of the relation is explicitly modeled, significantly improving the fit to high-$z$ data. However, the evidence for evolution strongly depends on how the likelihood function accounts for this intrinsic dispersion and is weaker if it is ignored. We also show that Malmquist bias significantly affects comparisons between low- and high-$z$ samples, reducing---though not eliminating---the statistical preference for redshift evolution after matching luminosity ranges. These results indicate that current HIIG data favor a redshift-dependent modification of the standard $L$--$\sigma$ relation, while highlighting the critical role of selection effects and intrinsic dispersion modeling in establishing HIIGs as precise cosmological probes.
\end{abstract}
\maketitle

\section{\label{sec:1}Introduction}
 
HIIGs are compact systems typically found in dwarf irregular galaxies undergoing intense bursts of star formation, driven by young super stellar clusters. Their optical spectra exhibit strong emission lines from gas ionized by these massive clusters within the host galaxies~\citep{1972ApJ...173...25S,1977ApJ...211...62B,1981MNRAS.195..839T,1987MNRAS.226..849M,1991A&AS...91..285T,2000A&ARv..10....1K,Bordalo:2011yc,Chavez:2014ria,Gonzalez-Moran:2021drc}.
A key feature of HIIGs is the empirical correlation between the integrated H$\beta$ luminosity ($L(\mathrm{H}\beta)$) and the ionized gas velocity dispersion ($\sigma$). This $L$--$\sigma$ relation enables HIIGs to serve as potential standard candles, offering a promising avenue for cosmological studies (e.g.,~\citep{1988MNRAS.235..297M,Siegel:2004xs,Plionis:2011jj,
Chavez:2012km,Terlevich:2015toa,FernandezArenas:2017dux,Gonzalez-Moran:2019uij,Gonzalez-Moran:2021drc,Chavez:2024twa}). 
The physical basis of the $L$--$\sigma$ relation lies in the scaling of both ionizing photon production ($\propto L(\mathrm{H}\beta)$) and gas kinematics ($\propto \sigma$) with the mass of the young stellar cluster~\citep{1981MNRAS.195..839T,1987MNRAS.226..849M,Bordalo:2011yc,Chavez:2014ria}. 

The use of HIIGs as ``standard candles'' heavily relies on the validity of $L$--$\sigma$ relation, and the empirical $L$--$\sigma$ relation has been widely discussed over the past few decades~\citep{1962IAUS...15..359S,1977ApJ...213...15M,1978A&A....70..157M,1979ApJ...228..394K,1981MNRAS.195..839T,1986A&A...156..111C,1987MNRAS.226..849M,2000MNRAS.311..629M,Bordalo:2011yc,Chavez:2014ria,2018MNRAS.474.4507L,Gonzalez-Moran:2021drc,Hernandez-Almada:2021rjs,Mehrabi:2021feg,2023PhRvD.107j3521C,2024MNRAS.527.7626R,2024PhRvD.109l3527C}. Notably,~\citet{Chavez:2014ria} compiled a sample of 128 HIIGs spanning a redshift range of $0.02 \lesssim z \lesssim 0.2$, where the integrated H$\beta$ fluxes (i.e., $f(\mathrm{H}\beta)$) are measured from low dispersion wide aperture spectrophotometry, and there is $f(\mathrm{H}\beta)\propto L(\mathrm{H}\beta)$; and the $\sigma$ are measured from the high equivalent widths of their Balmer emission lines with the observations of high S/N high-dispersion spectroscopy. Their findings demonstrate a strong and stable $L$--$\sigma$ relation within the selected sample.

While the local scaling $L$--$\sigma$ relation, i.e., $\log L(\mathrm{H}\beta)\propto \log \sigma$\footnote{where ``$\log$'' denotes the logarithm base 10.}, has been thoroughly characterized, its extrapolation to high-redshift regimes remains uncertain. The well-calibrated low-redshift relation may not necessarily hold for distant HIIGs~\citep{1995ApJ...440L..49K,1996ApJ...460L...5G,2000MNRAS.311..629M,2020ApJ...888..113W,2024PhRvD.109l3527C,2024ApJ...969...54W}.
Recent work by~\citet{2024PhRvD.109l3527C} has discovered redshift evolution in the $L$--$\sigma$ relation.
Their analysis reveals systematic variations in the relation's slope between local ($z < 0.2$) and distant ($z > 0.6$) populations, suggesting evolutionary effects that must be accounted for in cosmological applications.

The analysis of~\citet{2024PhRvD.109l3527C} used a joint fitting method, estimating parameters for both the $L$--$\sigma$ relation and a specific cosmological model. While this is a powerful approach, it directly links the properties of the astrophysical relation to the chosen cosmology. This mixing can introduce systematic errors into the inferred evolution, making the results dependent on the assumed cosmological model. To clearly separate the astrophysical signal from cosmology, we use a different, two-step method. First, we reconstruct the expansion history of the Universe in a way that does not assume a cosmology. We do this by applying Gaussian Process (GP) regression to the Pantheon+ sample of Type Ia supernovae (SNe Ia)~\citep{Brout:2022vxf} to get model-independent distance estimates up to $z \sim 1.8$. Using this reconstructed distance-redshift relation, we then perform a dedicated statistical comparison of different $L$--$\sigma$ relation models using only the HIIG data. This separate approach lets us: i) test for redshift evolution without imposing a cosmological model first, ii) carefully check how our conclusions depend on the distance reconstruction, on selection effects like Malmquist bias, and on how we model the data's intrinsic dispersion, and iii) present estimates for cosmological parameters based on different assumptions about the scaling relation and error model.

The paper is organized as follows: In Section~\ref{sec:data and gp}, we introduce the HIIG sample and explain how we reconstruct cosmology-independent distances using GP regression on the Pantheon+ SNe Ia data. Section~\ref{sec:Statistical Analysis and Model Comparison} presents a statistical model comparison between the classic $L$--$\sigma$ relation and several redshift-dependent versions. In Section~\ref{sec:Robustness Analysis}, we test how sensitive our conclusions are to the distance reconstruction, Malmquist bias, and the choice of statistical method. Section~\ref{sec:cosmology} provides cosmological parameter results under different assumptions for the scaling relation and error model, and examines how the parameters relate to each other. We summarize our findings in Section~\ref{sec:summary}.

\section{DATA AND METHODOLOGY}
\label{sec:data and gp}
\subsection{HIIGs dataset and empirical $L$--$\sigma$ relation}
\label{subsec:data}

Both giant extragalactic HII regions (GEHRs) and HIIGs are compact systems undergoing massive bursts of star formation~\citep{Chavez:2012km,1977ApJ...211...62B,2000A&ARv..10....1K,1981MNRAS.195..839T,1987MNRAS.226..849M,1988MNRAS.235..297M}. However, they differ in their host environments: GEHRs are typically found in the outer disks of late-type galaxies, whereas HIIGs reside in dwarf irregular galaxies. Due to their shared origin in intense star-forming activity, GEHRs and HIIGs exhibit nearly identical optical spectra, dominated by strong emission lines from gas ionized by young, massive star clusters. Consequently, both systems follow the $L$--$\sigma$ relation, though this relation primarily reflects the properties of the young starbursts rather than their host galaxies.
Given their spectral and dynamical similarities, GEHRs—being nearby—are often used as calibrators for the more distant HIIGs~\citep{1962IAUS...15..359S,1977ApJ...213...15M,1978A&A....70..157M,1979ApJ...228..394K,1981MNRAS.195..839T,1987MNRAS.226..849M,1988MNRAS.235..297M,Chavez:2012km,FernandezArenas:2017dux}.

The data set used in our analysis includes the measurements of 181 HIIGs and 36 GEHRs. The HIIGs sample comprises 107 low-redshift sources ($0.0088 < z < 0.1642$)~\citep{Chavez:2014ria}, and 74 high-redshift sources ($0.6364 < z < 2.5449$)~\citep{Terlevich:2015toa,Gonzalez-Moran:2019uij,Gonzalez-Moran:2021drc}. Our GEHRs sample originates from~\citet{FernandezArenas:2017dux} observations, where these GEHRs reside in 13 local ($z\sim0$) galaxies.

In this work, three steps are used to associate the observations of HIIGs or GEHRs with the cosmological distance, as described below.

\begin{enumerate}
\item [(i)] In the practical observations, one can obtain the integrated H$\beta$ emission-line flux, $f(\mathrm{H}\beta)$,  from wide-aperture low-resolution spectrophotometry, and measure the corresponding emission line velocity dispersion, $\sigma$, from high-resolution spectroscopy.

\item [(ii)] The luminosity, $L(\mathrm{H}\beta)$, for each HIIG or GEHR can be calculated based on a given $L$--$\sigma$ relation, e.g.,  the \textbf{classic scaling $L$--$\sigma$ relation}
\begin{equation}
	\log L{\rm{(H\beta)}} = \alpha + \beta \log \sigma.
	\label{eq:classls}
\end{equation}

\item [(iii)] Once obtaining $f(\rm{H}\beta)$ and $L(\rm{H}\beta)$ from the above steps and also using the definition of luminosity distance, i.e., $L = 4\pi  d_{\rm{L}}^2 f$, one can further calculate the luminosity distance ($d_{\rm{L}}$) for each HIIG or GEHR,
\begin{equation}
	d_{\rm{L}}^2 = \frac{L{\rm{(H\beta)}}}{4\pi f{\rm{(H\beta)}}}.
	\label{eq:f-l} 
\end{equation}
And then, the distance modulus $\mu(z)$ can be obtained as follows, 
\begin{equation}
	\mu(z) = 5\log d_{\rm{L}}(z)+25,
	\label{eq:dl-mu} 
\end{equation}
where $d_{\rm{L}}$ has units of $\rm{Mpc}$.
\end{enumerate}

\subsection{Reconstructing the Hubble diagram with SNe Ia data by using GP}
\label{subsec:gp}

As mentioned previously, to avoid the dependence on the cosmological model, we choose to reconstruct the Hubble diagram from SNe Ia data by employing the GP method.

The SNe Ia data used in this work is the  Pantheon+ sample, which includes 1701 light curves of 1550 distinct SNe Ia from 18 different sky surveys, with a redshift range of $0.001<z<2.26$~\citep{Brout:2022vxf}. In order to reduce systematics from peculiar velocities of nearby SNe Ia ($z<0.01$), we reconstruct the Hubble diagram using only SNe Ia in the redshift range $0.01<z<2.26$. The distance modulus is defined as
\begin{equation}
	\mu_{\mathrm{SN}}= m_{B} - M ,
	\label{eq:mu}
\end{equation}
where $m_{B}$ is the corrected apparent magnitude of the SNe Ia, which is calculated using the SALT2 method~\citep{PhysRevD.75.103508,SNLS:2010pgl,2017ApJ...836...56K,Brout:2022vxf,Brout:2021mpj}, and $M$ is the absolute magnitude. In this paper, the prior for the absolute magnitude of supernovae is set to the result calibrated by Cepheid variables as used by SH0ES, that is $M=-19.253\pm0.027$~\citep{Riess:2021jrx}.

\begin{figure*}[ht!]
\centering 
 \includegraphics[width=0.8\linewidth]{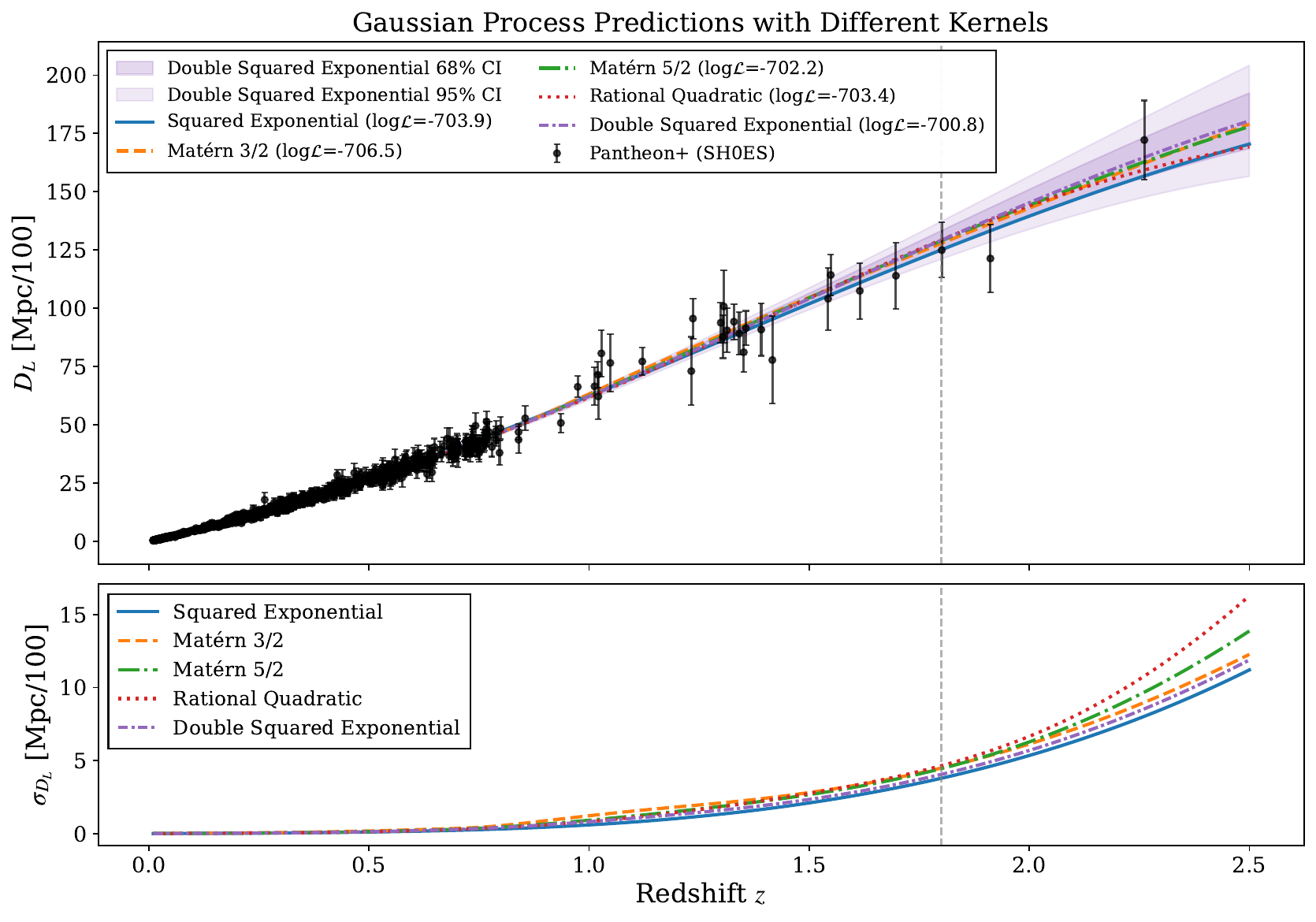}
	\caption{\textbf{The distance-redshift relation $D_\mathrm{L}(z)$ reconstructed via GP regression using the Pantheon+ sample of SNe Ia.} Here, $D_\mathrm{L}(z) = d_\mathrm{L}(z)$ [Mpc] / 100. The reconstruction adopts a Gaussian prior on the supernova absolute magnitude from SH0ES ($M = -19.253 \pm 0.027$). The top panel shows the Pantheon+ data and the mean predictions from five different GP kernels. The bottom panel displays the standard deviation (uncertainty) of the predictions from each kernel. The vertical dashed line marks a conservative redshift cut at $z_\mathrm{cut} = 1.8$, beyond which the supernova data become too sparse for a reliable reconstruction. Based on this cutoff, 145 of the 181 original HII galaxy observations are retained for our subsequent analysis. The plot demonstrates the consistency of the reconstructions across different kernels and the stability of the chosen double RBF kernel.}
    \label{fig:DL}
\end{figure*}

\begin{figure*}
	\centering
	\includegraphics[width=0.8\linewidth]{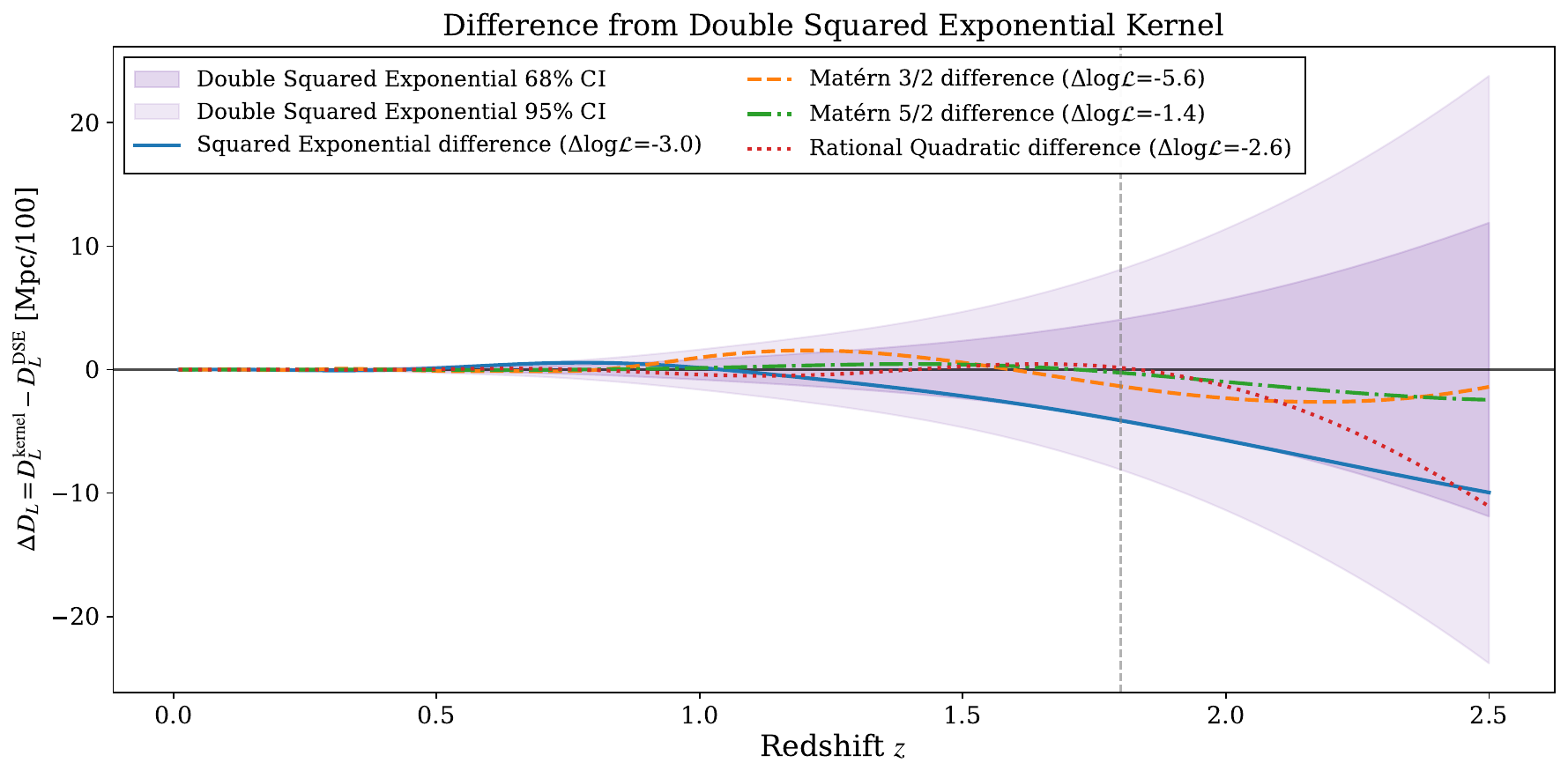}
	\caption{\textbf{Deviations of the GP-reconstructed distances from the Double RBF kernel prediction for different kernel choices.} The lines show the difference between the mean distance predicted by each of the four other kernels (Squared Exponential, Matérn~3/2, Matérn~5/2, Rational Quadratic) and the prediction from the Double RBF kernel. The shaded region represents the 1$\sigma$ uncertainty band (standard deviation) of the Double RBF reconstruction. The vertical dashed line indicates the conservative redshift cutoff at $z = 1.8$ adopted in our analysis to ensure reconstruction reliability.}
	\label{fig:delte_DL}
\end{figure*}

A GP generalizes the Gaussian distribution to functions of continuous variables. If a univariate function $f(x)$ follows a GP, then for any finite set of input points ${x_i}$, the corresponding function values ${f(x_i)}$ are jointly distributed as a multivariate Gaussian with a mean function $\mu(x)$ and a covariance (kernel) function $k(x, x')$ \citep{10.7551/mitpress/3206.001.0001}. This is denoted as:
\begin{equation}
f(x) \sim \mathcal{GP}\bigl(\mu(x), k(x, x')\bigr).
\label{eq:gp_def}
\end{equation}

The GP has been widely applied in cosmology~\citep{2012JCAP...06..036S,2012PhRvD..85l3530S,2022MNRAS.517..576H,2022ApJ...941...84L,2023PhRvD.108f3522Q,2024MNRAS.528.1354L}. 
Here we employ the open-source Python code \texttt{GaPP3}~\citep{2012JCAP...06..036S}, which determines the GP hyperparameters by maximizing the marginal likelihood and returns the posterior mean and variance of the reconstructed function.

To obtain cosmology-agnostic distance estimates from the Pantheon+ supernova sample, we perform GP regression. We tested several reconstruction targets: the distance modulus $\mu(z)$ directly, a zero-mean centered $\mu(z)$, and the luminosity distance $d_{\mathrm{L}}(z)$. Direct reconstruction of $\mu(z)$ exhibited unphysical oscillations at high redshift, while centering implicitly introduces a prior dependence on the chosen cosmological model. We therefore adopt the transformed luminosity distance $D_{\mathrm{L}}(z) = d_{\mathrm{L}}(z)[\mathrm{Mpc}] / 100$,
which behaves nearly linearly with redshift, promoting a stable and well-behaved GP reconstruction~\citep{Liao:2019qoc,Liao:2020zko,Liang:2022smf,2023JCAP...09..041M,Han:2024vkd,Li:2024hed,Rana:2025psr}. The covariance matrix for $D_{\mathrm{L}}$ is propagated from that of $\mu$ via standard error propagation:
\begin{equation}
\operatorname{Cov}(D_{\mathrm{L},i}, D_{\mathrm{L},j}) = \left(\frac{\ln 10}{5}\right)^2 D_{\mathrm{L},i} D_{\mathrm{L},j} \operatorname{Cov}(\mu_i, \mu_j).
\label{eq:cov_DL}
\end{equation}

For the reconstruction we set the mean function $\mu(z)$ to zero, ensuring no a priori cosmological model preference. An overall scale factor for the covariance matrix is also optimized to account for possible global misestimation of errors. To ensure robustness, we compare five commonly used kernels: the Squared Exponential (RBF), Double Squared Exponential (Double RBF), Matérn~3/2, Matérn~5/2, and Rational Quadratic kernels. The Double RBF kernel is particularly motivated for datasets such as Pantheon+ that combine uncertainties from distinct scales (e.g., intrinsic supernova dispersion and the cosmic distance ladder). Its functional form is
\begin{equation}
k(x, \tilde{x}) = \sigma_{f1}^2 \exp\!\left[-\frac{(x-\tilde{x})^2}{2\ell_1^2}\right]
                + \sigma_{f2}^2 \exp\!\left[-\frac{(x-\tilde{x})^2}{2\ell_2^2}\right],
\label{eq:double_rbf}
\end{equation}
where the four hyperparameters $(\sigma_{f1}, \ell_1, \sigma_{f2}, \ell_2)$ control the amplitude and correlation length of two components, allowing the kernel to capture processes operating on different characteristic scales.

Fig.~\ref{fig:DL} presents the reconstructed mean and variance obtained from all five kernel functions. To avoid convergence to local minima, we performed the hyperparameter optimization for each kernel from multiple starting points. All kernels yield consistent reconstructions of $D_{\mathrm{L}}(z)$, with no unphysical oscillations at high redshifts, and their predictive uncertainties are comparable. This indicates that the more flexible Double RBF kernel does not lead to overfitting. To ensure the reliability of the GP reconstruction, we apply a redshift cut at $z_\mathrm{cut}=1.8$, beyond which the supernova data become sparse. Based on this cutoff, we retain 145 of the 181 original HIIG observations for our subsequent analysis.

The stability of the reconstruction is further quantified in Fig.~\ref{fig:delte_DL}, which shows the difference between the mean prediction from each kernel and that of the Double RBF kernel. Over the relevant redshift range, all these differences lie well within the 95\% posterior uncertainty band of the Double RBF reconstruction. Kernels with similarly high marginal likelihoods (e.g., the Matérn~5/2 kernel) show only minor deviations. This confirms that the Double RBF kernel provides a reconstruction that is both stable and representative of the entire family of plausible smooth kernels.

After reconstructing the transformed luminosity distance $D_\mathrm{L}(z)$ through the GP, one can further calculate the luminosity $L(\mathrm{H}\beta)$ for each HIIG by using Eq.~(\ref{eq:f-l}). The uncertainty of luminosity is estimated with the following formula, 
\begin{equation}
	{\epsilon^2_{\log L}} = \frac{4}{25}{\epsilon^2_{\mu}}+{\epsilon^2_{\log f}},
	\label{eq:sigmalogl} 
\end{equation}
where ${\epsilon_{\mu}}$ and $\epsilon_{\log f}$ are the errors of the distance modulus $\mu(z)$ and the logarithm of emission line fluxes $\log f$, and ${\epsilon_{\mu}}$ is computed from Eq.~(\ref{eq:cov_DL}).

\section{Statistical Analysis and Model Comparison}
\label{sec:Statistical Analysis and Model Comparison}

\subsection{Corrections to the scaling $L$--$\sigma$ relation}
\label{subsec:corrections}
The analysis of~\citet{2024PhRvD.109l3527C} shows that the $L$--$\sigma$ relation for HIIGs is standardizable, however, there are significant differences between the slopes of the scaling $L$--$\sigma$ relation (i.e., Eq.~(\ref{eq:classls})) obtained from low-redshift and high-redshift subsamples, respectively. It suggests two possibilities: (i) the existence of the redshift evolution of the $L$--$\sigma$ relation; and (ii) the requirement of non-linear term in the $L$--$\sigma$ relation.

Based on the above mentioned possibilities, we propose three possible forms of corrections to the \textbf{classic scaling $L$--$\sigma$ relation}, i.e., Eq.~(\ref{eq:classls}), which are labeled as \textbf{Correction I--III}.

\begin{itemize}
\item \textbf{Correction I: Adding an extra redshift-evolutionary term}
\begin{equation}
	\log L{\mathrm{(H\beta)}} = \alpha + \beta \log \sigma + \gamma_1\log (1+\gamma_2 z).
    \label{eq:correction_1}
\end{equation}

\item \textbf{Correction II: Adding two logarithmic redshift-dependent coefficients}
\begin{equation}
\begin{aligned}
\log L({\mathrm{H\beta}}) = &[1+\gamma_1 \log (1+\frac{z}{1+z})] \alpha \\ &+ [1+\gamma_2\log (1+\frac{z}{1+z})] \beta \log \sigma.
\label{eq:correction_2}
\end{aligned}
\end{equation}

\item \textbf{Correction III: Adding two nonlinear redshift-dependent coefficients}
\begin{small}
\begin{equation}
\log L({\mathrm{H\beta}}) = (1+ \frac{\gamma_1 z}{1+z})\alpha + (1+\frac{\gamma_2 z}{1+z})\beta \log \sigma.
\label{eq:correction_3}
\end{equation}
\end{small}
\end{itemize}

\subsection{Statistical analysis}
\label{subsec:Statistical Analysis}

In our main analysis, the likelihood is constructed as
\begin{small}
\begin{equation}
	\begin{aligned}
		\mathcal{L} =  { \prod_{i=1}^{N}} \frac {1} {\sqrt{2\pi} \epsilon_{\mathrm{tot},i}} \times \exp \left[ -\frac{(\log L_{\mathrm{th},i}-\log L_{\mathrm{obs},i})^2}{2{\epsilon^2_{\mathrm{tot},i}}}  \right], 
		\label{eq:likelihood}
	\end{aligned}
\end{equation}
\end{small}
where $N$ is the number of data points, and the theoretical prediction of the luminosity $L_{\mathrm{th},i}$ is obtained from a certain $L$--$\sigma$ relation, where the four options for the $L$--$\sigma$ relation are presented in Eqs.~(\ref{eq:classls}) and (\ref{eq:correction_1})--(\ref{eq:correction_3}). For each HIIG, the observational value of the luminosity $L_{\mathrm{obs},i}$ can be obtained from Eqs.~(\ref{eq:f-l}) and (\ref{eq:dl-mu}) based on the $D_\mathrm{L}(z)$ reconstructed
through the GP; while for each GEHR, $L_{\mathrm{obs},i}$ is obtained from Eq.~(\ref{eq:f-l}) with the luminosity distance ($d_\mathrm{L}$) obtained from the local distance ladder.

In addition, $\epsilon_{\mathrm{tot},i}$ is the total uncertainty of $\log L_i$, the expression of which depends on the form of $L$--$\sigma$ relation. Specifically, in the case of choosing classic scaling form or \textbf{Correction I} for the $L$--$\sigma$ relation, the total uncertainty $\epsilon_{\mathrm{tot}}$ takes the following expression, 
\begin{equation}
	\epsilon^2_{\mathrm{tot}}=\epsilon^2_{\log L}+\beta^2{\epsilon^2_{\log \sigma}}+\epsilon^2_{\mathrm{int}}.
	\label{eq:classls_error}
\end{equation}
When choosing the \textbf{Correction II}, one obtains,
\begin{equation}
{\epsilon^2_{\mathrm{tot}}}={\epsilon^2_{\log L}}+[1+\gamma_2\log (1+\frac{z}{1+z})]^2\beta^2{\epsilon^2_{\log \sigma}}+\epsilon^2_{\mathrm{int}},
\label{eq:correction2_error}
\end{equation}
Correspondingly, in the scenario of \textbf{Correction III}, one has  
\begin{equation}
{\epsilon^2_{\mathrm{tot}}}={\epsilon^2_{\log L}}+(1+\frac{\gamma_2 z}{1+z})^2\beta^2{\epsilon^2_{\log \sigma}}+\epsilon^2_{\mathrm{int}}.
\label{eq:bilinearls_error}
\end{equation}
Furthermore, in Eqs.~(\ref{eq:classls_error})--(\ref{eq:bilinearls_error}), $\epsilon^2_{\log L}$ is calculated with Eq.~(\ref{eq:sigmalogl}), $\epsilon_{\log\sigma}$ is the uncertainty of the logarithm of emission line velocity dispersion $\sigma$, which is obtained from the actual observations, and $\epsilon_{\mathrm{int}}$ serves as a measure of the intrinsic dispersion in $\log L$.

Following~\citet{2024ApJ...964L...4C}, we compute the posterior probability distributions for the model parameters and the Bayesian evidence by using the Python open-source package \texttt{PyMultiNest}~\citep{Buchner:2014nha}, which serves as an interface to the \texttt{MultiNest} algorithm~\citep{10.1111/j.1365-2966.2009.14548.x} based on nested sampling~\citep{Skilling:2004pqw}. The \texttt{GetDist}~\citep{2019arXiv191013970L} is used to plot the marginalized 1-D and 2-D posterior probability distributions for the parameters.

\subsection{Model Comparison for The $L$--$\sigma$ Relation}
\label{subsec:Model Comparison}
Observational constraints for the $L$--$\sigma$ relation parameters are summarized in the upper portion of Table~\ref{table:sh0es}, where a joint sample with 36 GEHRs and 145 HIIGs\footnote{As discussed in Section \ref{subsec:gp}, the reconstruction for transformed luminosity distance $D_\mathrm{L}(z)$ from the GP is more credible in the range of $z \lesssim 1.8$, and this redshift range covers 145 HIIGs among the entire 181. Here we need to use the reconstructed $D_\mathrm{L}(z)$ for the HIIGs, so only 145 HIIGs are used.} is employed, and four options for the $L$--$\sigma$ relation, including the classic scaling form in Eq.~(\ref{eq:classls}) and the three correction forms in Eqs.~(\ref{eq:correction_1})--(\ref{eq:correction_3}), are taken into account, respectively.

We compute the posterior probability distributions for the model parameters using the \texttt{PyMultiNest} code, which implements the nested sampling algorithm via the \texttt{MultiNest} library. From the resulting weighted posterior samples, we extract the median value for each parameter as the central estimate and construct credible intervals (CI) based on posterior quantiles. The median values and their 68\% CI are reported in Table~\ref{table:sh0es}. In this work, both the 68\% and 95\% CI are defined directly from the cumulative posterior distribution of the weighted samples. For subsequent statistical comparisons--such as evaluating parameter differences between models or samples--the relevant probabilities (e.g., $P(\theta_i > \theta_j)$) are also computed directly from the weights of the posterior samples.

To compare the three proposed modifications of the $L$--$\sigma$ relation against the classic scaling form, we employ the Bayesian evidence as the model selection criterion~\citep{Kass:1995loi,Trotta:2008qt}. For each model we compute the natural logarithm of the Bayesian evidence, $\ln \mathrm{B}_i$, and the relative log-Bayesian evidence $\ln \mathrm{B}_{i0} = \ln \mathrm{B}_i-\ln \mathrm{B}_0$, where $\ln \mathrm{B}_0$ corresponds to the classic scaling relation. These values, obtained using the \texttt{PyMultiNest} code, are listed in the final two columns of Table~\ref{table:sh0es}. Interpretation follows the empirical Jeffreys scale: values of $\ln \mathrm{B}_{i0}$ in the ranges $(0,1.0)$, $(1.0,2.5)$, $(2.5,5.0)$, and $(>5.0)$ indicate inconclusive, weak, moderate, and strong evidence in favor of model $i$ over the reference model, respectively~\citep{Trotta:2008qt}. A positive and sufficiently large $\ln \mathrm{B}_{i0}$ therefore signals statistical preference for a given modified relation.

\begin{table*}[htbp]
        \renewcommand\arraystretch{1.5}
        \setlength{\tabcolsep}{6pt}
        \footnotesize
        \centering
	\caption{Parameter estimates for the HII galaxy $L$--$\sigma$ relation from the combined GEHR and HIIG sample}\label{table:sh0es}
		\begin{tabular}{ccccccccc}
			\hline
Likelihood functions
& $L$--$\sigma$ relation & $\alpha$ & $\beta$ & $\gamma_1$ & $\gamma_2$  &$\epsilon_{\mathrm{int}}$  & $\ln \mathrm{B}_i$ & $\ln \mathrm{B}_{i0}$  \\
			\hline
			& Classic relation & $34.13^{+0.20}_{-0.19}$& $4.42^{+0.13}_{-0.13}$& $\dots$&$\dots$&$0.29^{+0.02}_{-0.02}$   & $-73.61$ & $0$ \\
			\cline{2-9}
\textbf{Full $\mathcal{L} $ } 			& \textbf{Correction I}& $34.09^{+0.21}_{-0.19}$& $4.45^{+0.13}_{-0.15}$ $\star$& $-0.03^{+0.71}_{-0.59}$ $\star$&$0.28^{+2.44}_{-0.33}$ $\star$&$0.29^{+0.02}_{-0.02}$& $-76.46$ & $-2.90$ \\
			\cline{2-9}
\textbf{with $\epsilon_{\rm{int}} = \mathrm{Const.}$}			& \textbf{Correction II}& $33.79^{+0.19}_{-0.20}$& $4.64^{+0.14}_{-0.13}$&  $0.92^{+0.10}_{-0.11}$&$-3.97^{+0.45}_{-0.42}$&$0.26^{+0.02}_{-0.02}$& $-57.30$ & $16.02$ \\
			\cline{2-9}
			& \textbf{Correction III}& $33.79^{+0.19}_{-0.19}$& $4.65^{+0.12}_{-0.13}$&  $0.31^{+0.04}_{-0.04}$&$-1.32^{+0.15}_{-0.14}$ & $0.26^{+0.02}_{-0.02}$&$-60.31$ & $13.28$ \\
			\hline
                \hline
& Classic relation & $33.48^{+0.08}_{-0.08}$& $4.85^{+0.06}_{-0.05}$& $\dots$&$\dots$ &$\dots$  & $-217.37$ & $0$ \\
			\cline{2-9}
\textbf{Full $\mathcal{L} $ } &	\textbf{Correction I}& $33.47^{+0.09}_{-0.09}$& $4.86^{+0.06}_{-0.06}$& $0.01^{+0.51}_{-0.50}$ &$0.16^{+2.65}_{-0.19}$ $\star$&$\dots$& $-220.34$ & $-2.97$ \\
			\cline{2-9}
\textbf{with $\epsilon_{\rm{int}} = \mathrm{0.}$}&	\textbf{Correction II}& $33.46^{+0.09}_{-0.10}$& $4.87^{+0.07}_{-0.07}$&  $0.42^{+0.09}_{-0.10}$&$-1.70^{+0.39}_{-0.35}$&$\dots$& $-219.27$ & $-1.90$ \\
			\cline{2-9}
		&	\textbf{Correction III}& $33.45^{+0.09}_{-0.10}$& $4.87^{+0.07}_{-0.07}$&  $0.14^{+0.03}_{-0.04}$ $\star$&$-0.56^{+0.15}_{-0.12}$ $\star$& $\dots$&$-221.86$ & $-4.46$ \\
                \hline
                \hline
& Classic relation & $33.36^{+0.08}_{-0.08}$& $4.93^{+0.05}_{-0.06}$& $\dots$&$\dots$ &$\dots$ & $-358.72$ & $0$ \\
			\cline{2-9}
\textbf{$\mathcal{L} \propto \mathrm{exp}(-\mathrm{\chi}^2/2)$} &	\textbf{Correction I}& $33.33^{+0.09}_{-0.09}$& $4.96^{+0.07}_{-0.06}$& $-0.06^{+0.46}_{-1.02} \star$ &$0.16^{+2.47}_{-0.19} \star$ &$\dots$& $-361.40$ & $-2.68$ \\
			\cline{2-9}
&	\textbf{Correction II}& $33.34^{+0.09}_{-0.10}$& $4.96^{+0.07}_{-0.06}$&  $-0.14^{+0.16}_{-0.18}$&$0.52^{+0.71}_{-0.63}$&$\dots$& $-366.31$ & $-7.59$ \\
			\cline{2-9}
		&	\textbf{Correction III}& $33.34^{+0.09}_{-0.10}$& $4.96^{+0.07}_{-0.07}$&  $-0.06^{+0.06}_{-0.06}$&$0.25^{+0.24}_{-0.23}$& $\dots$&$-367.15$ & $-8.43$ \\
			\hline
            
	  \end{tabular}
      
\footnote{Observational constraints on the classic $L$--$\sigma$ relation and three modified versions (\textbf{Corrections I--III}) from a combined sample of 36 GEHRs and 145 HIIGs. The table reports the median values and 68\% CI from the parameter posterior distributions, computed using three different likelihood functions. When a posterior distribution is noticeably asymmetric (typically indicated by a difference between the upper and lower bounds exceeding 15\% of the median value), the corresponding entry is marked with an asterisk ($\star$). A direct check of the contour plots confirms these entries are consistent with the visual results. These asymmetric posteriors primarily involve parameters that are non-essential or poorly constrained. For each model, we also list the log-Bayesian evidence ($\ln \mathrm{B}_i$) and the relative evidence compared to the classic model ($\ln \mathrm{B}_{i0} \equiv \ln \mathrm{B}_i - \ln \mathrm{B}_0$).}
\end{table*}

For \textbf{Correction I}, the 68\% CI for the correction parameters $(\gamma_1,\gamma_2)$ include $(0, 0)$, showing that the classic relation remains compatible with the data. The negative value $\ln \mathrm{B}_{10} = -2.90$ ($2.5<|\ln \mathrm{B}_{10}|<5.0$) provides moderate evidence in favor of the classic relation over \textbf{Correction I}. In contrast, for \textbf{Correction II} the correction parameters satisfy $\gamma_1 > 0$ and $\gamma_2 < 0$ at the 99\% CI. The large positive evidence $\ln \mathrm{B}_{20} = 16.02$ ($|\ln \mathrm{B}_{20}| > 5.0$) constitutes strong evidence supporting this model. Similarly, for \textbf{Correction III} we find $\gamma_1 > 0$ and $\gamma_2 < 0$ at the 99\% CI, with $\ln \mathrm{B}_{30} = 13.28$ also indicating strong evidence. Since $|\ln \mathrm{B}_{20}| > |\ln \mathrm{B}_{30}|$, \textbf{Correction II} is the most favored among the three modified parameterizations.

Overall, \textbf{Correction II} is statistically preferred, indicating significant evidence for redshift evolution in the $L$--$\sigma$ relation within this phenomenological framework. We note, however, that this redshift parameterization likely acts as a proxy for unobserved physical properties (e.g., size or metallicity of the star-forming region~\citep{Chavez:2014ria}). The current scarcity of such ancillary data for high-redshift HIIGs highlights the need for future observations, more physically motivated models, and alternative analysis approaches.

\begin{figure*}[ht!]
\centering 
 \includegraphics[width=0.7\linewidth]{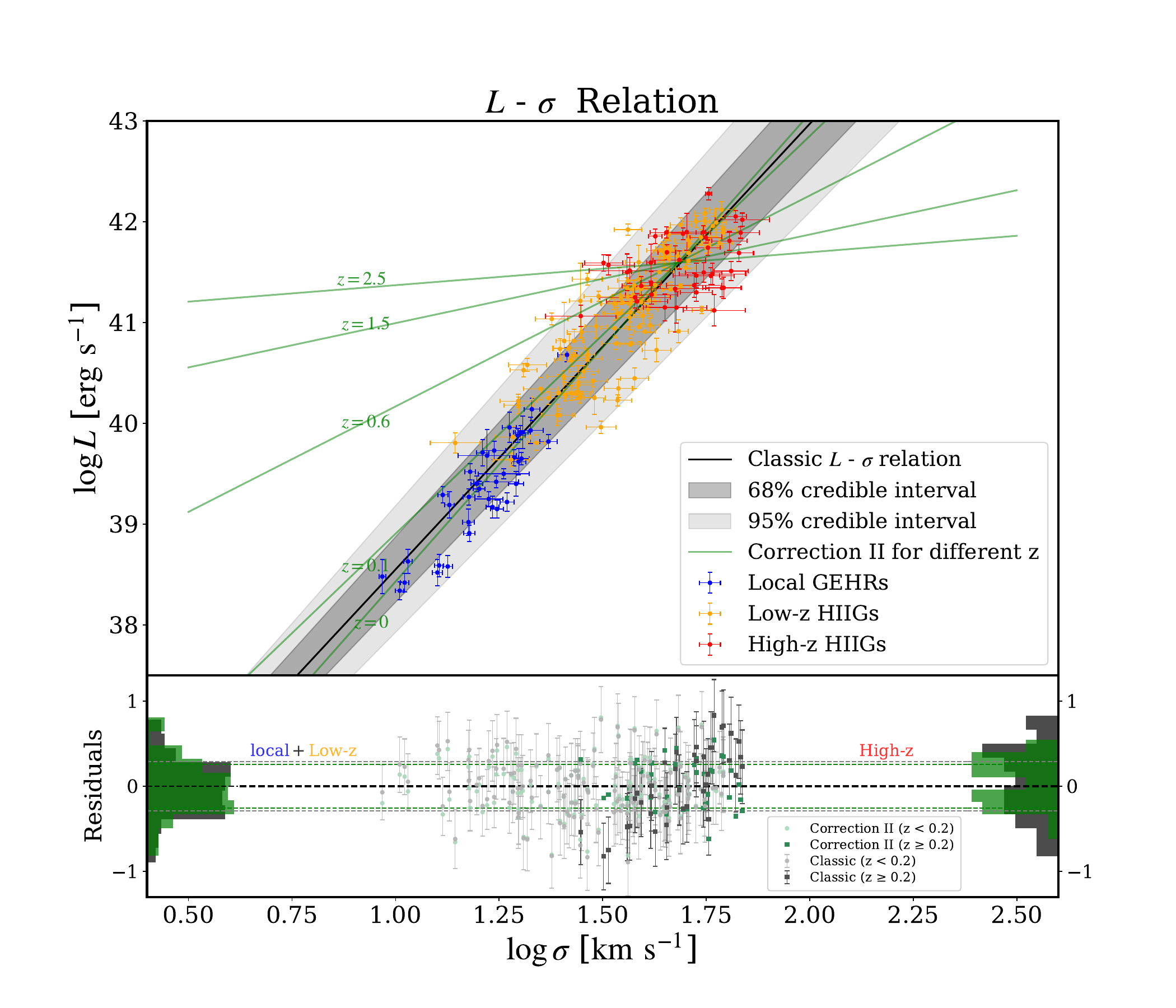}
	\caption{\textbf{Comparison between the classic and a modified $L$--$\sigma$ relation for HIIGs.}
\textbf{Top panel:} The black solid line and shaded bands show the classic relation (Eq.~(\ref{eq:classls})) using the median parameter values and their 68\% and 95\% CI from Table~\ref{table:sh0es}. The green lines illustrate the modified relation (\textbf{Correction II}, Eq.~(\ref{eq:correction_2})) evaluated at different redshifts (from $z=0$ to $2.5$), using the corresponding median parameters.
\textbf{Bottom panel:} Residuals in $\log L$ relative to the predictions of the classic relation (black) and the modified relation (green). Histograms separate the local + low-z objects (GEHRs and HIIGs with $z<0.2$) from the high-z HIIGs ($z>0.6$). The modified relation removes the systematic trend seen in the classic residuals at high redshift.}
    \label{fig:classic-correction2}
\end{figure*}

Fig.~\ref{fig:classic-correction2} provides a visual comparison between the \textbf{Correction II} parameterization and the classic $L$--$\sigma$ relation, showing their posterior predictions against the data along with a residual analysis. Theoretical values are derived from the posterior medians of the parameters (Table~\ref{table:sh0es}), with shaded bands indicating the 68\% and 95\% CI for the classic model.

The upper panel displays the theoretical predictions. The classic $L$--$\sigma$ relation, while supported by the majority of the data and providing relatively precise predictions, shows a potential systematic issue: its 68\% credible band is not aligned with the principal axis of the high-redshift data distribution. For \textbf{Correction II}, we plot a family of curves corresponding to redshifts from $0$ to $2.5$ to illustrate the direction and magnitude of the evolution. These curves better track the main trend of the high-redshift data. The predictive intervals for \textbf{Correction II} are significantly broader at high redshifts; for clarity, these bands are not shown in the figure, but the substantial predictive uncertainty they imply is addressed in the following discussion.

The lower panel presents the residuals of $\log L$ with respect to the predictions of each model. For the classic relation, while the high-redshift residuals cluster near zero, they exhibit a systematic, slanted pattern in the $\log\sigma$ space. This indicates that the classic model fails to capture a redshift-dependent trend correlated with velocity dispersion. In contrast, the residuals for \textbf{Correction II} are centered on zero and appear randomly scattered for both low- and high-redshift data, demonstrating that its redshift-dependent term successfully absorbs this systematic trend.

In summary, the classic model achieves higher fitting precision for the local calibrators but exhibits a clear systematic bias for high-redshift objects. \textbf{Correction II} eliminates this bias at the cost of introducing significantly larger predictive uncertainty, particularly at high redshifts. The Bayesian model comparison naturally quantifies this trade-off: the substantially larger marginal likelihood (Table~\ref{table:sh0es}) for \textbf{Correction II} indicates that its more accurate, unbiased description of the data outweighs the penalty from its increased parametric freedom and predictive variance. Therefore, the joint analysis of the figure and the Bayesian evidence supports the existence of redshift evolution in the $L$--$\sigma$ relation, with the form of \textbf{Correction II} being statistically preferred. Nevertheless, the larger predictive uncertainty of \textbf{Correction II} at high redshifts also implies its reduced effectiveness for making precise cosmological predictions.

\section{Robustness Analysis: Testing Key Assumptions}
\label{sec:Robustness Analysis}

The posterior probability distributions of the model parameters and the Bayesian evidence together provide statistically significant support for redshift evolution within the examined parameterizations of the $L$--$\sigma$ relation, with the \textbf{Correction II} scenario being the most strongly favored.

We now examine the robustness of this result by assessing potential systematic influences on the inferred evolutionary signal. Specifically, we investigate three key aspects that could impact the observed trend: (1) the prior on the SNe Ia absolute magnitude used in the distance reconstruction, (2) the effects of Malmquist bias arising from sample selection, and (3) the formulation of the likelihood function in the HIIG data analysis.

\subsection{Evaluating the Sensitivity to SNe Ia Absolute Magnitude Priors}
\label{subsec:Mb}

As discussed in Section~\ref{subsec:gp}, we must specify a prior on the absolute magnitude $M$ of SNe Ia when using GP to reconstruct the Hubble diagram from SNe Ia data. While our baseline analysis adopts the SH0ES prior ($M = -19.253 \pm 0.027$), we also test an alternative prior from the Multicolor Light Curve Shape (MLCS) method ($M = -19.33 \pm 0.25$)~\citep{2000ApJ...536..531W}. 

The negligible impact on our key findings suggests that the results are not sensitive to the specific choice of $M$ prior, reinforcing their robustness.

\subsection{Assessing the Impact of Malmquist Bias}
\label{subsec:malmquist bias}
The observed high-redshift HIIGs are dominated by the most luminous systems due to selection effects, specifically Malmquist bias. To assess its impact, we compare the posterior distributions under the classic $L$--$\sigma$ relation between a high-$z$ subsample ($z > 0.2$, $\log L > 41$) and a luminosity-matched low-$z$ subsample ($z < 0.2$, $\log L > 41$). We perform a Bayesian inference by directly constructing the posterior samples of the parameter differences $\Delta\theta = \theta_{\mathrm{low-z}} - \theta_{\mathrm{high-z}}$.

For the matched samples, the posterior median difference is $\Delta\alpha = -2.23$ with a 95\% CI of $[-3.93, -0.39]$, and $\Delta\beta = 1.36$ with a 95\% CI of $[0.26, 2.38]$. The corresponding posterior probabilities are $P(\Delta\alpha < 0) = 99.1\%$ and $P(\Delta\beta > 0) = 99.2\%$, indicating strong evidence for parameter differences between the redshift regimes.

If the low-$z$ sample is not luminosity-matched, the evidence for parameter differences becomes substantially stronger: we find $\Delta\alpha = -4.65$ with a 95\% CI of $[-6.18, -3.01]$ and $P(\Delta\alpha < 0) > 99.9\%$, and $\Delta\beta = 2.76$ with a 95\% CI of $[1.77, 3.67]$ and $P(\Delta\beta > 0) > 99.9\%$. Applying a traditional Gaussian approximation to the point estimates yields significances of approximately $2.5\sigma$ and $5.5\sigma$ for the matched and unmatched samples, respectively, consistent with the full posterior analysis.

We adopt a simplified subsample comparison rather than weighting the data by a specific selection function, as the HIIG dataset is compiled from different instruments; imposing a uniform selection model could introduce additional systematics. Future studies with a detailed understanding of each instrument should re-examine methods for direct Malmquist bias correction prior to $L$--$\sigma$ relation analysis.

In summary, the Bayesian posterior probability analysis indicates that even after a basic luminosity-matching correction for Malmquist bias, evidence for redshift evolution in the $L$--$\sigma$ relation persists at high significance (probability $>95\%$). However, the discrepancy between high- and low-redshift samples is greatly reduced after this correction. This suggests that selection effects are a significant, though not the sole, contributor to the observed differences.

\subsection{Testing the Influence of the Likelihood Function}
\label{subsec:Likelihood Function}

In the main analysis of this work, we adopt the likelihood function given by Eq.~(\ref{eq:likelihood}). In this form, the total error includes a free parameter, $\epsilon_\mathrm{int}$, which adaptively accounts for the intrinsic dispersion and various systematic uncertainties in the scaling relation. The theoretical foundation for this form can be found in~\citet{DAgostini:2005mth}. It has been widely used in the development of standard candles, including gamma-ray bursts (GRBs) and quasars~\citep{Risaliti:2015zla,Khadka:2020hvb,Lusso:2020pdb,Liu:2022srx,Han:2024vkd,Li:2024hed,Lusso:2025bhy,Wu:2025pmx}, as well as in previous studies of the HIIG scaling relation~\citep{2024PhRvD.109l3527C}.

Given the current SALT2 method for SNe Ia~\citep{PhysRevD.75.103508,SNLS:2010pgl,2017ApJ...836...56K,Brout:2022vxf,Brout:2021mpj} and studies of the low-redshift HIIG scaling relation~\citep{Chavez:2014ria}, one expects the HIIG relation to require corrections from other observables that may evolve with redshift. Ideally, each data point would have its own correction term. However, current observational precision and the lack of strong prior knowledge make a hierarchical Bayesian approach with per-datum parameters impractical. Thus, we approximate the combined dispersion and systematic errors as a single isotropic term, $\epsilon_\mathrm{int}$, which is an acceptable simplification.

Nevertheless, because allowing $\epsilon_\mathrm{int}$ to vary freely is an approximate error model (the true individual $\epsilon_{\mathrm{int},i}$ being unknown), we also test a likelihood with $\epsilon_\mathrm{int}=0$, which corresponds to neglecting intrinsic dispersion and unknown systematic errors.

For comparison, we also consider a likelihood commonly employed in HIIG cosmological parameter estimation~\citep{Gonzalez-Moran:2019uij,Gonzalez-Moran:2021drc}. This form also sets $\epsilon_\mathrm{int}=0$ and omits the normalization factor. Although Bayesian posterior estimation should always include the normalization factor, we examine this likelihood due to its widespread use in cosmological parameter estimation.

The three likelihood forms are labeled as follows:
\begin{itemize}
\item \textbf{Full $\mathcal{L} $ with $\epsilon_{\rm{int}} = \mathrm{Const.}$}

In this case, the likelihood function is expressed as Eq.~(\ref{eq:likelihood}), with the intrinsic dispersion $\epsilon_{\rm{int}}$ treated as a free parameter.

\item \textbf{Full $\mathcal{L}$ with $\epsilon_{\rm{int}} =0$}

In this case, the likelihood function is given by Eq.~(\ref{eq:likelihood}), where the intrinsic dispersion $\epsilon_{\rm{int}}$ is fixed at zero.

\item \textbf{$\mathcal{L} \propto \mathrm{exp}(-\mathrm{\chi}^2/2)$}

In this case, the likelihood function is expressed as
\begin{equation}
\mathcal{L}\propto e^{-\chi^2/2},
\label{eq:likelihood_chi2}
\end{equation}
where $\chi^2$ is
\begin{equation}
\chi^2 = \sum_{i=1}^{N}\frac{(\log L_{\mathrm{th,i}}-\log L_{\mathrm{obs,i}})^2}{{\epsilon^2_{\mathrm{tot,i}}}} ,
\label{eq:chi2}
\end{equation}
and $N$ is the number of data points.

\end{itemize}

The median values and 68\% CI of the posterior distributions for the classic $L$--$\sigma$ relation and the three modified models, derived from the three likelihood functions, are also presented in Table~\ref{table:sh0es}.

Based on the results in Table~\ref{table:sh0es}, we find that when using \textbf{Full $\mathcal{L}$ with $\epsilon_{\rm{int}} =0$}, \textbf{Correction II} still yields the highest evidence value among the three correction models, and the parameters $(\gamma_1, \gamma_2)$ continue to exclude zero at the 99\% CI. This indicates persistent signs of redshift evolution in the scaling relation. However, according to the Bayesian evidence under this likelihood formulation, the evidence now favors the classic relation over the correction models. We speculate that when employing \textbf{Full $\mathcal{L}$ with $\epsilon_{\rm{int}} =0$}, the local and low-redshift data---which have smaller observational errors---are given relatively greater statistical weight, while the influence of the higher-error, high-redshift data is diminished. This shifts the posterior medians for the redshift-evolution parameters closer to zero. Meanwhile, the reduction in the number of free parameters increases the posterior precision for the remaining parameters, which still allows zero to be excluded within relatively large credible interval. Correspondingly, the strong Bayesian evidence for redshift evolution under \textbf{Full $\mathcal{L}$ with $\epsilon_{\rm{int}} = \mathrm{Const.}$} stems primarily from a superior fit to the high-redshift data. The reduced weighting of these data under \textbf{Full $\mathcal{L}$ with $\epsilon_{\rm{int}} =0$} naturally diminishes the Bayesian evidence in favor of \textbf{Correction II}.

The likelihood form \textbf{$\mathcal{L} \propto \exp(-\chi^2/2)$} exhibits a distinct behavior and notable stability. Under all three correction models, the 68\% CI for the correction parameters derived from this likelihood consistently include zero. Furthermore, this simplified form has been used historically for joint calibration of the scaling relation and cosmological parameter estimation with reasonable success. We conjecture that this arises from a peculiar balance between omitting an explicit model for $\epsilon_\mathrm{int}$ and neglecting the normalization factor in the likelihood, coupled with the implicit assumption of no redshift evolution during parameter estimation. These simplifications may inadvertently mitigate several inherent issues in the data modeling, though this interpretation currently lacks definitive evidence.

In summary, while we maintain a preference for \textbf{Full $\mathcal{L}$ with $\epsilon_{\rm{int}} = \mathrm{Const.}$} based on its explicit and physically motivated treatment of intrinsic dispersion, our comparative analysis demonstrates that the inferred evidence for redshift evolution is closely tied to the specific choice of likelihood function. We therefore recommend that future studies place greater emphasis on the detailed characterization of error sources in HIIG data and, crucially, on investigating the physical origin of the intrinsic dispersion. As with other emerging standard candles, this is a complex challenge that will require both a deeper understanding of the fundamental physics governing the $L$--$\sigma$ relation and larger datasets to robustly constrain its properties.

\section{Cosmological Application}
\label{sec:cosmology}
We assess the cosmological constraints from the combined HIIG and GEHR sample. Specifically, we constrain the $\Lambda$CDM model using 36 GEHRs and all 181 HIIGs, applying three likelihood functions to both the classic $L$--$\sigma$ relation and the most competitive modified form, \textbf{Correction II}.

\begin{figure}[htbp]
    \centering
    \includegraphics[scale=0.57]{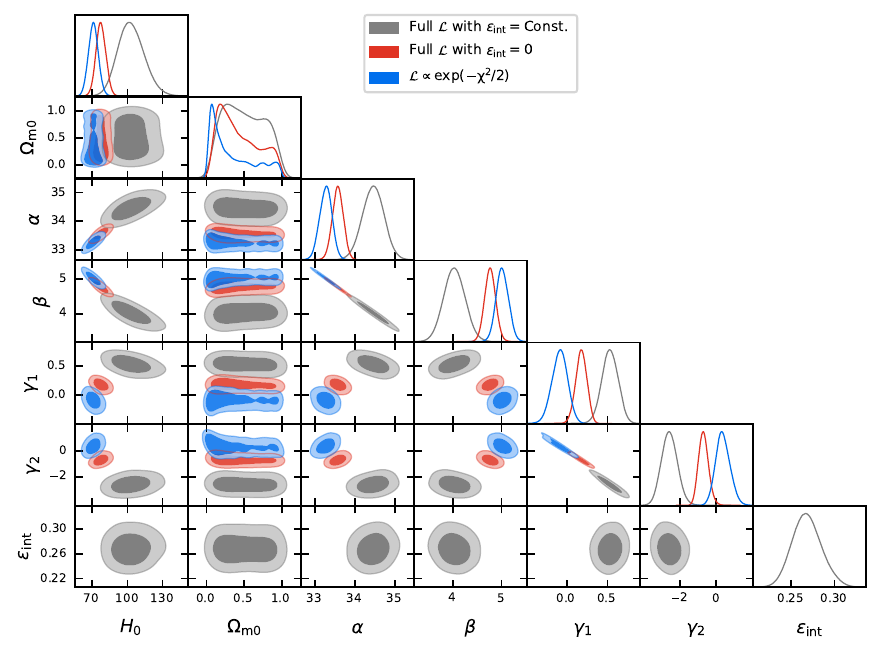}
    \caption{Cosmological parameter constraints ($H_0$,$\Omega_{m0}$) within the $\Lambda$CDM model, derived from the combined GEHR and HIIG sample using the redshift-dependent \textbf{Correction II} for the $L$--$\sigma$ relation. The contours show results from three different statistical likelihoods.}
    \label{fig:lcdm_c2}
\end{figure}

The median values and 68\% CI for all parameters are presented in Table~\ref{table:cosmology_correction2}. For the classic relation, the posteriors are approximately Gaussian, whereas for \textbf{Correction II} the matter density parameter $\Omega_{m0}$ exhibits a non-Gaussian distribution. The marginalized one- and two-dimensional posteriors under \textbf{Correction II} are shown in Fig.~\ref{fig:lcdm_c2} to facilitate the following discussion.

First, we present the constraints on the cosmological parameters $H_0$ and $\Omega_{m0}$. Using the classic $L$--$\sigma$ relation, the 68\% CI for $H_0$ are $126.03^{+12.32}_{-11.29}$, $81.38^{+4.35}_{-4.60}$, and $69.39^{+4.11}_{-4.12}$ for the three likelihoods (full $\mathcal{L}$ with $\epsilon{\rm{int}} = \mathrm{Const.}$, full $\mathcal{L}$ with $\epsilon_{\rm{int}} = 0$, and $\mathcal{L} \propto \exp(-\chi^2/2)$, respectively). The corresponding $\Omega_{m0}$ constraints are $0.68^{+0.17}_{-0.16}$, $0.42^{+0.07}_{-0.06}$, and $0.30^{+0.05}_{-0.05}$. We compare these to the Planck 2018 base-$\Lambda$CDM results ($H_0 = 67.4 \pm 0.5$, $\Omega{m0} = 0.315 \pm 0.007$; \citep{Planck2018}) by computing the posterior probability $P(\theta_{\mathrm{HIIG}} > \theta_{\mathrm{Planck}})$. For $H_0$, these probabilities are $>99.9\%$, $> 99.9\%$, and $68.3\%$; for $\Omega_{m0}$, they are $99.7\%$, $97.3\%$, and $41.4\%$, respectively. This highlights a strong dependence on the likelihood choice. Only the simplified $\mathcal{L} \propto \exp(-\chi^2/2)$ yields $H_0$ consistent with Planck within 68\% CI. The full likelihoods, particularly with $\epsilon_{\rm{int}} = \mathrm{Const.}$, produce $H_0$ values significantly higher (outside the 95\% CI) than the Planck value.

For \textbf{Correction II}, the tension with Planck is reduced. The corresponding probabilities become $P(\Delta_{H_0}>0)>99.9\%$, $98.8\%$, and $78.2\%$, and $P(\Delta_{\Omega_{m0}}>0)=68.8\%$, $56.1\%$, and $40.4\%$ for the three likelihoods. While \textbf{Correction II} lowers the median $H_0$ for the two full likelihoods, it does not significantly resolve the tension. For $\Omega_{m0}$, zero difference with Planck is now within the 68\% CI for all likelihoods, though the median estimates are also lowered. A key trade-off is evident: the constraining power on $\Omega_{m0}$ is drastically weakened under \textbf{Correction II}, consistent with its rapidly growing predictive uncertainty at high redshift noted in Subsection \ref{subsec:Model Comparison}. This limits its current practicality for precise cosmological forecasting.

Next, we examine the constraints on the $L$--$\sigma$ relation parameters within \textbf{Correction II}. For the full $\mathcal{L}$ with $\epsilon_{\rm{int}} = \mathrm{Const.}$, we find $(\alpha, \beta, \gamma_1, \gamma_2) =$ $(34.48^{+0.25}_{-0.27}$, $4.02^{+0.21}_{-0.20}$, $0.52^{+0.10}_{-0.09}$, $-2.56^{+0.44}_{-0.43})$. For the full $\mathcal{L}$ with $\epsilon_{\rm{int}} = 0$, the values are $(33.57^{+0.15}_{-0.14}$, $4.76^{+0.12}_{-0.12}$, $0.18^{+0.07}_{-0.07}$, $-0.70^{+0.28}_{-0.27})$. For $\mathcal{L} \propto \exp(-\chi^2/2)$, they are $(33.27^{+0.16}_{-0.16}$, $5.01^{+0.13}_{-0.13}$, $-0.08^{+0.09}_{-0.11}$, $0.35^{+0.46}_{-0.38})$. In the first two cases, $(\gamma_1, \gamma_2)$ exclude (0,0) at 95\% CI, indicating persistent signs of evolution. However, under $\mathcal{L} \propto \exp(-\chi^2/2)$, (0,0) lies within the 68\% CI. These differences stem from complex parameter degeneracies, illustrated in Fig.~\ref{fig:lcdm_c2}. Key correlations include positive trends in the $H_0$--$\alpha$, $H_0$--$\gamma_2$, $\alpha$--$\gamma_2$, and $\beta$--$\gamma_1$ planes, and negative trends in the $H_{0}$--$\beta$, $H_{0}$--$\gamma_1$, $\alpha$--$\gamma_1$, $\beta$--$\gamma_2$, and $\gamma_1$--$\gamma_2$ planes.

Analyzing these degeneracies clarifies the impact of the scaling relation on cosmology. Under the classic relation, a larger $\alpha$ yields a larger $H_0$, while a larger $\beta$ produces a smaller $\Omega_{m0}$. \textbf{Correction II} inherits this structure but adds couplings: a larger $\gamma_1$ reduces $\alpha$ (thus lowering $H_0$), and a smaller (more negative) $\gamma_2$ increases $\beta$ (thus lowering $\Omega_{m0}$). This explains how the evolutionary model shifts the median cosmological estimates.

The stark differences between likelihoods are also traceable through these degeneracies. Fig.~\ref{fig:lcdm_c2} shows a mild degeneracy between $\epsilon_\mathrm{int}$ and $(\alpha, \beta, \gamma_1, \gamma_2)$. Fixing $\epsilon_\mathrm{int}=0$ forces $\alpha$ smaller, $\beta$ larger, $\gamma_1$ smaller, and $\gamma_2$ larger, which through the scaling-cosmology linkages results in lower $H_0$ and $\Omega_{m0}$ estimates—exactly as observed. Furthermore, omitting the normalization factor in $\mathcal{L} \propto \exp(-\chi^2/2)$ biases $\beta$ upward and $\alpha$ downward at maximum likelihood, further affecting the cosmological constraints.

In summary, the constraints on $H_0$ and $\Omega_m$ from GEHRs+HIIGs are approximately an order of magnitude less precise than those from Planck 2018. When redshift evolution is accounted for via \textbf{Correction II}, the constraint on $\Omega_{m0}$ becomes particularly weak, undermining the utility of high-redshift HIIGs. Only the simplified likelihood $\mathcal{L} \propto \exp(-\chi^2/2)$ yields agreement with Planck within 68\% CI. The specific likelihood formulation significantly influences the inference of redshift evolution, which in turn directly affects cosmological parameter estimates through parameter degeneracies linked to intrinsic dispersion modeling. Future progress requires a deeper physical understanding of the $L$--$\sigma$ relation's intrinsic dispersion, or larger samples of high-redshift HIIGs with ancillary data, to refine the error model and clarify the source of dispersion.

\begin{table*}[htbp]
        \renewcommand\arraystretch{1.5}
        \setlength{\tabcolsep}{6pt}
        \footnotesize
        \centering
	\caption{Joint constraints on cosmological and  $L$--$\sigma$ scaling relation parameters (68\% CI).}\label{table:cosmology_correction2}
		\begin{tabular}{ccccccccc}
			\hline
			Likelihood functions& $L$--$\sigma$ relation&$H_0$ & $\Omega_{m0}$ &$\alpha$ & $\beta$ & $\gamma_1$ & $\gamma_2$  &$\epsilon_{\mathrm{int}}$   \\
			\hline
			Full $\mathcal{L} $& classic relation&$126.03^{+12.32}_{-11.29}$ & $0.68^{+0.17}_{-0.16}$ & $35.03^{+0.23}_{-0.24}$ &$3.57^{+0.19}_{-0.18}$ &$
            \dots$  & $\dots$ & $0.28^{+0.02}_{-0.02}$ \\
			\cline{2-9}
			 with $\epsilon_{\rm{int}}=\mathrm{Const.}$ &\textbf{Correction II}& $103.16^{+11.46}_{-10.28}$ & $0.47^{+0.34}_{-0.27}$ & $34.48^{+0.25}_{-0.27}$ &$4.02^{+0.21}_{-0.20}$ &$0.52^{+0.10}_{-0.09}$  & $-2.56^{+0.44}_{-0.43}$ & $0.27^{+0.02}_{-0.02}$ \\
			\hline
			\hline
			Full $\mathcal{L} $& classic relation&$81.38^{+4.35}_{-4.60}$& $0.42^{+0.07}_{-0.06}$& $33.68^{+0.13}_{-0.13}$ &$4.68^{+0.11}_{-0.11}$ &$
            \dots$  & $\dots$ & $(0.00)$ \\
			\cline{2-9}
			with $\epsilon_{\rm{int}}=0$& \textbf{Correction II}&$77.24^{+4.48}_{-4.13}$& $0.37^{+0.39}_{-0.23}$& $33.57^{+0.15}_{-0.14}$ &$4.76^{+0.12}_{-0.12}$ &$0.18^{+0.07}_{-0.07}$& $-0.70^{+0.28}_{-0.27}$ & $(0.00)$ \\
            \hline
            \hline
			$\mathcal{L} \propto$&classic relation& $69.39^{+4.11}_{-4.12}$& $0.30^{+0.05}_{-0.05}$& $33.21^{+0.14}_{-0.15}$ &$5.06^{+0.12}_{-0.12}$ &$\dots$& $\dots$ & $\dots$ \\
			\cline{2-9}
			$\mathrm{exp}(-\mathrm{\chi}^2/2)$&\textbf{Correction II}&$70.87^{+4.41}_{-4.41}$& $0.21^{+0.52}_{-0.16}$& $33.27^{+0.16}_{-0.16}$ &$ 5.01^{+0.13}_{-0.13}$ &$-0.08^{+0.09}_{-0.11}$& $0.35^{+0.46}_{-0.38}$ & $\dots$ \\
            \hline
	  \end{tabular}
      
\footnote{Cosmological parameter estimates for the $\Lambda$CDM model, using 36 GEHR and 181 HII galaxy measurements. Results are shown for both the classic and the modified (\textbf{Correction II}) $L$--$\sigma$ relations, analyzed with three different likelihood functions.}
\end{table*}

\section{Summary}
\label{sec:summary}

We present a cosmology-independent study of the $L$--$\sigma$ relation in HIIGs to test its reliability as a standard candle and look for possible changes with redshift.

We use GP reconstruction of the distance-redshift relation from the Pantheon+ SNe Ia data. This lets us compare four different forms of the $L$--$\sigma$ relation without assuming a specific cosmological model. We find clear signs that the classic, non-evolving relation does not fully describe the data. A redshift-dependent correction is needed. Among the models we test, \textbf{Correction II}--which adds a term linking $\log \sigma$ to $\log(1+z/(1+z))$--is strongly preferred by the Bayesian evidence. This model successfully removes systematic trends in the high-redshift data, leaving random residuals (Fig.~\ref{fig:classic-correction2}). This supports the idea that the $L$--$\sigma$ relation evolves with redshift.

We check the strength of this result. Our findings are robust against the choice of supernova absolute magnitude prior and the details of the GP distance reconstruction. We also test the impact of Malmquist bias by comparing luminosity-matched samples at low and high redshift. While this selection effect reduces the statistical preference for evolution, significant differences between redshift bins remain (posterior probabilities $>95\%$).

However, there are important caveats. The evidence for evolution depends heavily on how we model the intrinsic dispersion in the data. The preference for a redshift-dependent correction is strongest when we use a full likelihood that includes a free intrinsic dispersion parameter ($\epsilon_{\mathrm{int}}$). This preference becomes much weaker with simpler likelihoods. Also, \textbf{Correction II} is a practical, not physical, model. It likely captures the effect of unmeasured galaxy properties (like metallicity or size). While it fits the data better, it also leads to much larger uncertainties in predicting luminosities at high redshift. This weakens the resulting cosmological constraints (Table~\ref{table:cosmology_correction2}), limiting its current use for precise cosmology.

In conclusion, to make HIIGs a reliable precision tool for cosmology, we need a better physical understanding of the $L$--$\sigma$ relation and the source of its intrinsic dispersion. Future work should focus on getting larger, high-quality samples of high-redshift HIIGs, along with measurements of other properties. This is key to telling real redshift evolution apart from selection effects and moving beyond simple fitting formulas.

\begin{acknowledgments}
 We are deeply grateful to Dr.~Roberto Terlevich for insightful discussions on investigating whether the observed redshift evolution in the $L$--$\sigma$ relation reflects a genuine physical effect. We also thank the anonymous referees for their time and valuable feedback, which helped improve the manuscript. This work has been supported by the National Key Research and Development Program of China (No.~2022YFA1602903) and the National Natural Science Foundation of China (Nos.~12588202, 12473002, 12075042, and 12475047).
\end{acknowledgments}

\nocite{*}

\bibliographystyle{apsrev4-2}
\bibliography{new}
\clearpage

\end{document}